\documentclass[12pt,a4paper]{article}
\usepackage{graphicx,fancyhdr}
\usepackage{cite}
\usepackage{color}
\begin{document}

\pagenumbering{arabic}

\vspace{-40mm}

\begin{center}
\Large\textbf{Magnetoresistance of granular Pt-C nanostructures close to the metal-insulator-transition}\normalsize \vspace{5mm}

\large F. Porrati, R. Sachser and M. Huth

\vspace{5mm}

\small\textit{Physikalisches Institut, Goethe-Universit\"at, Max-von-Laue-Str.~1, D-60438 Frankfurt am Main, Germany}

\end{center}

\vspace{0cm}
\begin{center}
\large\textbf{Abstract}\normalsize
\end{center}

We investigate the electrical and magneto-transport properties of Pt-C granular metals prepared by focused-electron-beam induced deposition. In particular, we consider samples close to the metal-insulator-transition obtained from as-grown deposits by means of a low-energy electron irradiation treatment. The temperature dependence of the conductivity shows a $\sigma\sim lnT$ behavior with a transition to $\sigma\sim \sqrt{T}$ at low temperature, as expected for systems in the strong-coupling tunneling regime. The magnetoresistance is positive and is described within the wave-function shrinkage model, normally used for disordered system in the weak-coupling regime. In order to fit the experimental data spin-dependent tunneling has to be taken into account. In the discussion we attribute the origin of the spin-dependency to confinement effects of Pt nano-grains embedded in the carbon matrix.

\newpage

\begin{center}
\large\textbf{1. Introduction}\normalsize
\end{center}

Granular metals are disordered electronic systems made of metallic nanoparticles embedded in a dielectric matrix. In the last decade, they attracted growing interest in the field of nanoscience and nanotechnology~\cite{beloborodov,huth}. This is driven by the technical ability to control the matter at the nanoscale and, thus, to the possibility to fabricate new materials with tunable electrical, magnetic and optical properties, as by focused electron beam induced deposition (FEBID) which is a direct writing technique used to fabricate granular metal nanostructures~\cite{utke,huth}. In FEBID the electron beam of a scanning electron microscope (SEM) is focused on a substrate in order to dissociate the adsorbate molecules of a precursor gas previously introduced in the SEM chamber. The solid part of the fragmentation reaction forms the sample, which is obtained by rastering the beam over the area of interest.

The electrical transport properties of granular metals are governed by the electron tunneling between neighboring grains through the dielectric matrix~\cite{beloborodov}. Depending on the value of the intergrain coupling $\textit{g}$, the electrical transport regime is either insulating ($g\leq1$, weak coupling) or metallic ($g\geq1$, strong coupling). In the weak coupling regime granular FEBID samples often show a stretched-exponential behavior in the temperature dependence of the conductivity, the so called correlated variable range hopping (correlated-VRH)~\cite{huth,beloborodov2005,feigelman}, which was first derived by Efros-Shklowski for disordered systems in the presence of Coulomb interaction (ES-VRH)~\cite{efros}. In correlated-VRH, since the Coulomb blockade tends to suppress the electron tunneling between neighboring grains at low bias voltages, higher order tunneling processes (co-tunneling) are involved in the electrical transport. In the strong coupling regime, the Coulomb interaction is screened and the electrons tunnel easily (quasi-metallic tunneling regime). In this regime, the theory predicts a logarithmic temperature dependence of the conductivity with a crossover to $\sqrt{T}$ at low temperatures for the three dimensional case~\cite{beloborodov2003}. This behavior was recently measured in Pt-C nanostructures fabricated by FEBID~\cite{sachser}. Note that the available models for the electrical transport in granular metals are applicable either in the weak or in the strong tunneling regime but not in the region of the metal-insulator-transition (MIT)~\cite{beloborodov}.

Recently the magnetotransport properties of granular metals have been investigated theoretically. At very low temperatures, in the correlated-VRH regime, elastic co-tunneling dominates the electrical transport~\cite{tran}. In this regime in the presence of a magnetic field interference effects among various paths connecting two grain sites lead to a negative magnetoresistance~\cite{feigelman}. In the inelastic co-tunneling regime, on the other hand, the interference is destroyed and the magnetoresistance is expected to be zero~\cite{feigelman}. In the strong coupling regime it was shown that the application of a magnetic field easily suppresses the localization correction (negative magnetoresistance)~\cite{beloborodov2004,biagini,blanter}, as known for disordered systems~\cite{kawabata}.

In this paper, we present electrical and magneto-transport measurements of granular Pt-C nanostructures prepared by FEBID. We chose FEBID as preparation technique for the granular metal samples, because it allows for a fine tuning of the intergrain coupling strength by post-growth irradiation~\cite{sachser,fab2011,schwalb}. In particular, we focus on the regime close to the MIT where available transport models are not applicable~\cite{beloborodov}. We find that the magnetoresistance is always positive and that it depends on the temperature and on the irradiation time. In order to interpret the data we tentatively extend the use of the wave function shrinkage model to the region of the MIT. By taking into account a spin-dependent electron tunneling contribution, we obtain excellent fits to the experimental data. Accordingly, we conclude that Pt nano-grains of FEBID nanostructures are magnetic, as also reported for Au, Pd and Pt nanoparticles in other works~\cite{yamamoto2004,shinohara,yamamoto2003,liu}.

\begin{center}
\large\textbf{2. Experimental details}\normalsize
\end{center}

The samples were prepared by using a dual-beam SEM/FIB microscope (FEI, Nova Nanolab 600) equipped with a Schottky electron emitter. The metal-organic precursor gas $(CH_3)_3CH_3C_5H_4Pt$ was injected in the SEM by a capillary with 0.5~mm inner diameter in close proximity to the focus of the electron beam on the substrate surface. The samples were grown on (100) Si(p-doped)/SiO$_2$ (200~nm) substrates. Contact electrodes of 60~nm Au/Cr were prepared by UV lithography. Samples $\#$1 and $\#$2 ($\#$3) were prepared by repeatedly rastering the area of interest with the electron beam with energy, current, pitch and dwell time of 5~keV, 0.1~nA (1~nA), 10~nm (20~nm) and 100~$\mu$s (1~$\mu$s), respectively. The composition of the deposits was about 23$\%$~Pt and 77$\%$~C independently of the beam parameters used during growth, as measured by energy dispersive x-ray analysis (EDX). The deposits had an area of $0.5\times5~\mu m^2$. The thickness of the samples after postgrowth electron irradiation treatment was about 125~nm for sample$\#$1 and 40~nm for samples $\#$2 and $\#$3, as revealed by AFM measurements (Nanosurf, easyscan2).

The electrical and magneto-transport measurements were performed in the temperature range from 1.8 to 265~K in two different variable-temperature insert ${^4}$He cryostats equipped either with a 9~T or with a 12~T superconducting solenoid. In particular, two-probe measurements were carried out with a Keithley Sourcemeter 2400 by fixing either the voltage for temperature-dependent conductivity measurements or the current for magnetoresistance measurements. In independent measurements we verified that the influence of the contact resistances on the resistance values was negligible~\cite{huth2009}.

\begin{center}
\large\textbf{3. Results}\normalsize
\end{center}

\begin{figure}\vspace{-18cm}\center{\includegraphics[width=22cm]{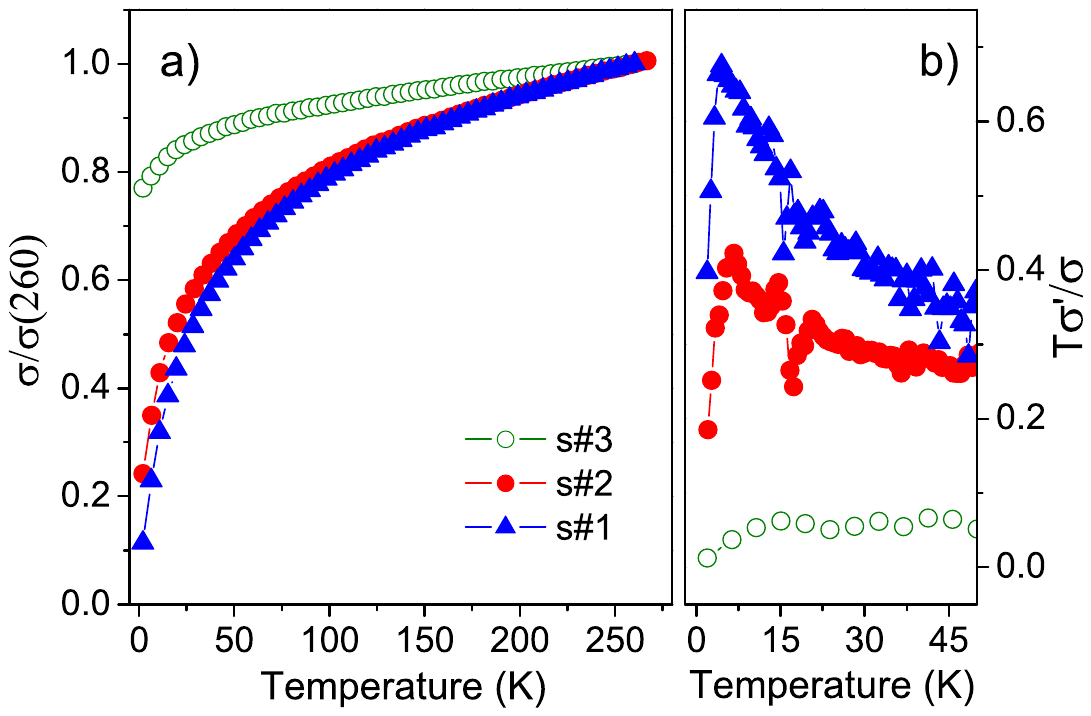}}
\caption{Left: Temperature dependence of conductivity for three deposits treated  by post-growth electron irradiation. Thickness of the deposits: Sample $\#$1 125~nm, after 0.32~$\mu C/\mu m^2$ irradiation dose; sample $\#$2 and $\#$3, ca. 40~nm after 0.32~$\mu C/\mu m^2$ and 1.16~$\mu C/\mu m^2$ dose, respectively. The irradiation was carried out with 5~keV beam energy and 1~nm electron beam current. Right: Logarithmic derivative of the conductivity suggesting that all the samples are on the metallic side of the MIT.}
\label{conductivity}\end{figure}

\begin{figure}\vspace{-18cm}\center{\includegraphics[width=22cm]{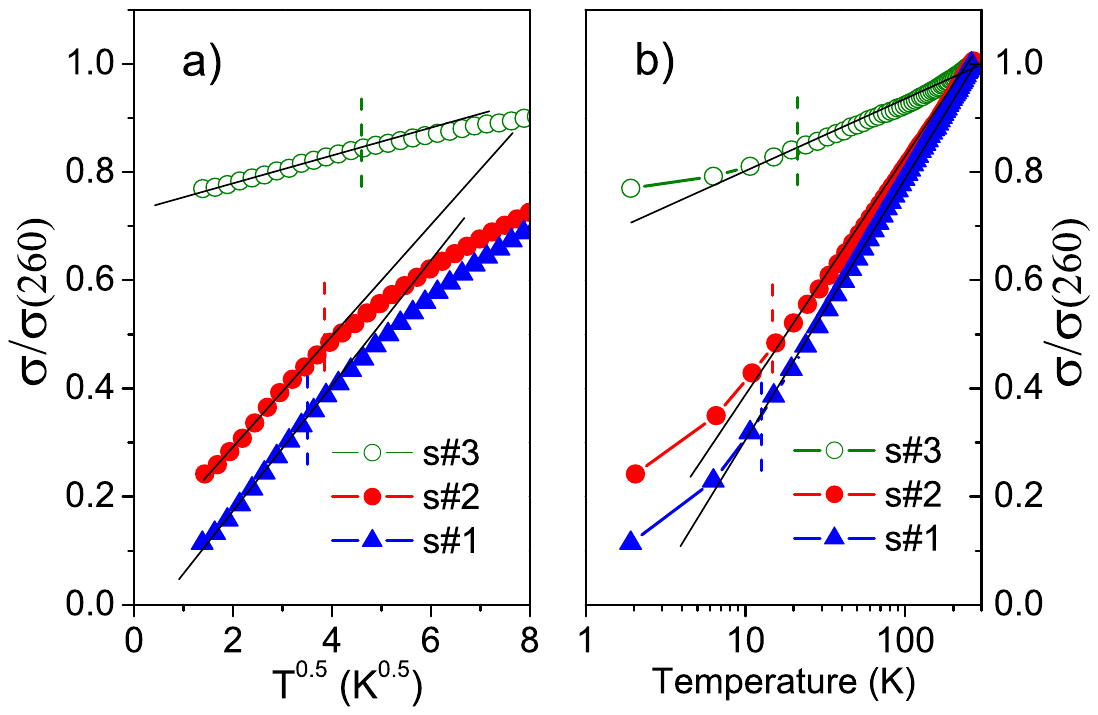}}
\caption{Conductivity vs. temperature of granular Pt-C samples $\#$1, $\#$2 and $\#$3. a) At low temperatures the electrical conductivity follows a $\sqrt{T}$ behavior. b) Above 20~K the conductivity shows a logarithmic temperature dependence. The crossover temperatures, which are marked by dashed lines, are T=12.2~K, T=14.7~K, and T=21~K, for samples $\#$1, $\#$2 and $\#$3, respectively.}
\label{conductivity_2}\end{figure}

In Fig.~\ref{conductivity} we plot the conductivity vs. temperature of the three samples used in this work. As-grown Pt-C FEBID samples are insulating and show a temperature dependence of the conductivity typical for correlated-VRH~\cite{fab2011,sachser}. The behavior of the conductivity for the postgrowth irradiated samples is typical of quasi-metallic samples, with finite conductivity at low temperature (Fig.~\ref{conductivity}a). All three samples are on the metallic side of the metal-insulator-transition, as indicated by the logarithmic derivative of the conductivity $T\cdot~\sigma^\prime/\sigma$ which tends to zero for $T\rightarrow0$, see Fig.~\ref{conductivity}b~\cite{moebius}. The conductivity increases with the irradiation dose, as seen by comparing samples $\#$2 and $\#$3. This is due to the transformation of the matrix microstructure which follows a graphitization trajectory between amorphous carbon and nanocrystalline graphite, to the slight enlargement of the Pt nanograins and to the decrease of the peripheral distance between neighboring grains~\cite{fab2011}. Furthermore, from Fig.~\ref{conductivity}a one can notice that the conductivity of sample $\#$1 is slightly smaller than that of sample $\#$2, although the irradiation dose is equal. This is due to the different thicknesses of the samples. In particular, since the penetration depth of 5~keV electrons in Pt-C is about 100~nm, the irradiation affects the complete volume of sample $\#$2, while this is not the case for the thicker sample $\#$1. In Fig.~\ref{conductivity_2} we analyze in detail the dependence of the electrical conductivity on the temperature. Below 20~K the conductivity follows a $\sqrt{T}$ behavior, while above 20~K it shows a logarithmic temperature dependence. Such a crossover between two conductance regimes was predicted about a decade ago~\cite{beloborodov2003} and recently measured in similar Pt-C FEBID deposits~\cite{sachser}.

\begin{figure}\vspace{-18cm}\center{\includegraphics[width=22cm]{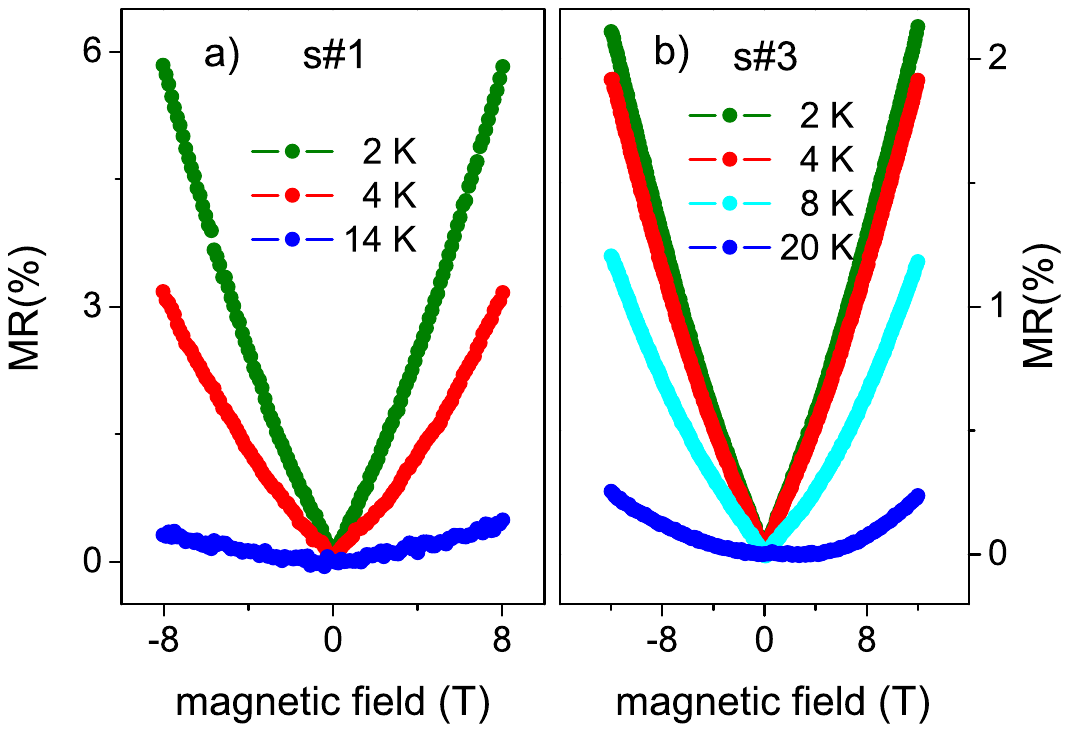}}
\caption{a) Large-field overview of magnetoresistance of sample $\#$1 and b) sample $\#$3. The magnetoresistance increases with decreasing temperature or for larger irradiation time, i.e., enhanced coupling strength.}
\label{MR}\end{figure}

\begin{figure}\vspace{-18cm}\center{\includegraphics[width=22cm]{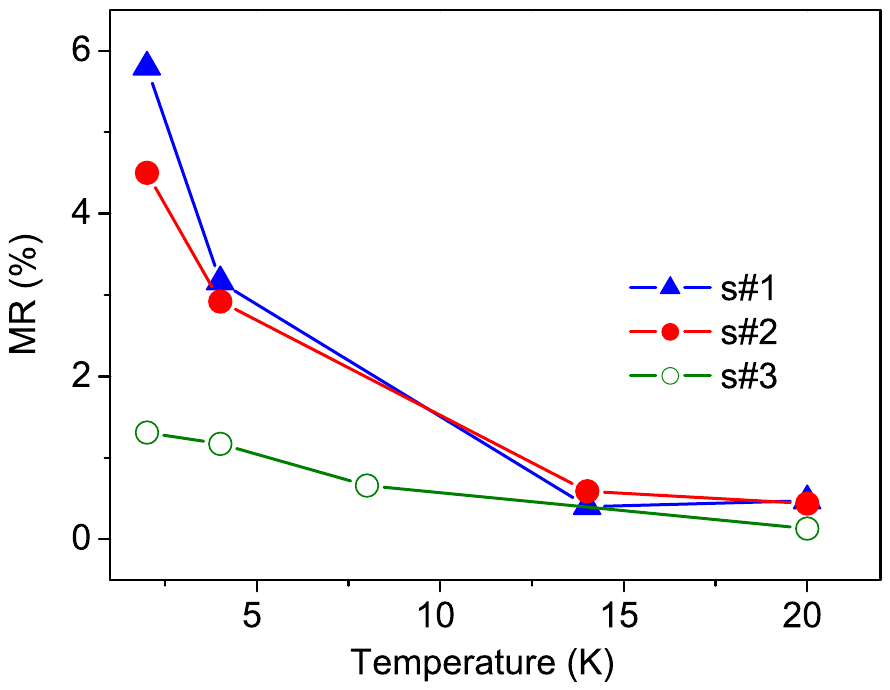}}
\caption{Temperature dependence of the MR at 8~T for samples $\#$1$, $\#$2$ and \#$3$. The MR decrease with the increase of temperature and intergrain coupling strength.}
\label{MR_2}\end{figure}

In Fig.~\ref{MR} we plot the magnetoresistance (MR) as function of the temperature for samples $\#$1 and $\#$3. The MR is always positive and increases at lower temperatures. Above ca.~20~K the MR is not detectable anymore. At the same constant temperature, the  MR of sample $\#$1 is larger than that of $\#$3, see also Fig.~\ref{MR_2}. This clearly indicates that the MR increases by moving towards the metal-insulator-transition from the metallic side. For clean Pt thin films, the MR is one or two orders of magnitude smaller than the one measured here\cite{hoffmann}. On the insulating side of the MIT, i.e., in the correlated-VRH regime, the MR is expected to be zero (inelastic co-tunneling)~\cite{feigelman}. In our experiment, attempts to detect the MR in this regime were not successful, which may be attributed either to a low signal-to-noise ratio or to a zero MR.

\begin{figure}\vspace{-18cm}\center{\includegraphics[width=24cm]{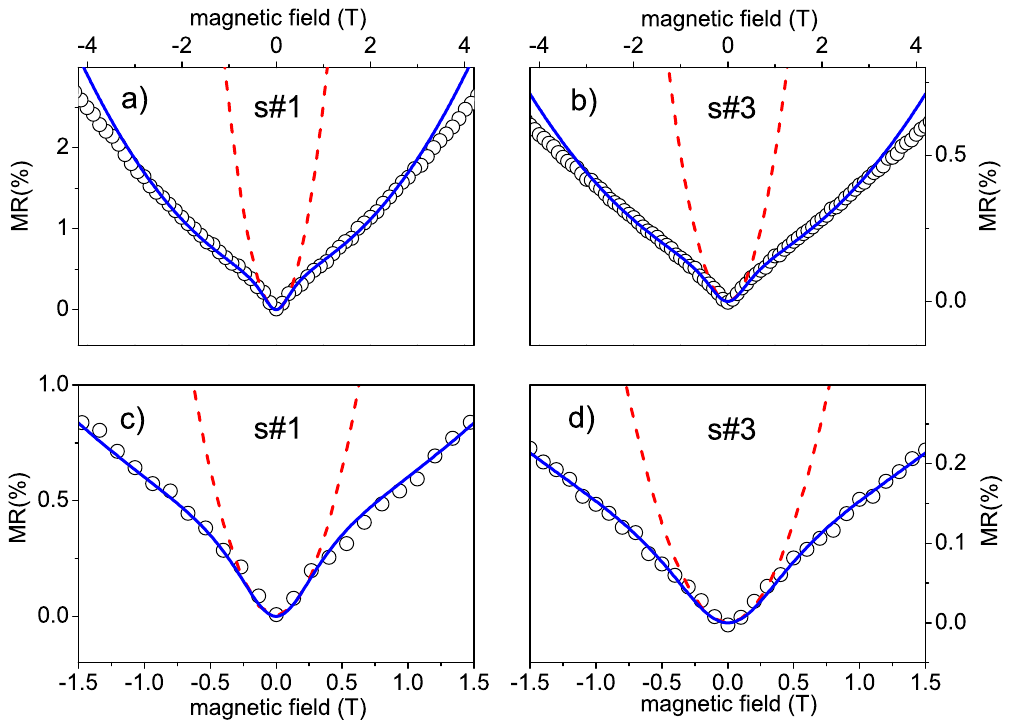}}
\caption{Magnetoresistance of samples~\#1 and \#3  at 2 K. Red curves: fits obtained by using equation (3) with $P_{i,j}$=0 and m=0. The fits deviate from the experimental data for $H\geq0.3$~T. Blue curves: fits obtained by using equation (3) with $P_{i,j}\neq0$ and $m\neq0$, i.e., by considering the Pt nano-grains as magnetic. The fit parameters used are shown in Fig.~\ref{parameters}. The fits deviate from the experimental data for $H\geq2.5$~T. All the fit are performed with the support of a gaussian weight function with standard deviation equal to the magnetic field corresponding to the second inflection point, at about 1~T in the figure.}
\label{fit_MR}\end{figure}

In order to interpret the MR data we tentatively use the $\textit{wave function shrinkage model}$, which was developed originally for bulk semiconductor systems in the VRH regime~\cite{shklovskii}. We are led to apply this model out of its theoretical predicted validity range, i.e., in the insulating regime, due to a peculiar low-field feature of the MR (see Fig.~\ref{fit_MR}) which is discussed next.

Within the wave function shrinkage model a wave function for an electron tunneling between two grain sites $i$ and $j$ of the following form is assumed:

\begin{equation}
\psi(r_{ij})\propto~exp(-\frac{r_{ij}}{\xi}-f(H))
\label{equation_1}\end{equation}

where $r_{ij}$ is the distance between the grains, $\xi$ is the localization length and $f(H)$ is a function of the applied magnetic field. The presence of a magnetic field reduces the overlap of neighboring electron wave functions and, thus, the electron tunneling probability, leading to a positive magnetoresistance~\cite{efros}. In particular, an extension of the model to grains with finite spin polarization was suggested in Refs.~\cite{boff2002,boff2013} which we found to be of relevance in the present case. For the extended model the resistance is given by:

\begin{equation}
R(H,T)=\frac{R(0,T)}{1+<P_iP_j>m^2}~exp(AH^2)
\label{equation_2}\end{equation}

with $R(0,T)=R_0~exp(T_0/T)^{1/2}$ the resistance in the ES-VRH regime, with $T_0$ a characteristic temperature, $P_i$ and $P_j$ the spin-polarization of the grains $i$ and $j$, $m=coth(\mu H/k_BT)-k_BT/\mu H$ the Langevin function with $\mu$ the magnetic moment of the grains, $A$ a parameter equal to $\frac{e^2\xi^4}{\alpha\hbar^2}(T_0/T)^{3/2}$~\cite{schoepe} with $e$ the elementary charge, $\alpha$ a numerical factor, and $\hbar$ the reduced constant of Planck. The results of the simple model~\cite{efros,schoepe} are recovered for $P_{i,j}=0$ and $m=0$. The correlation function $<P_iP_j>$ depends on the electronic density of states (DOS), the applied bias voltage and the hopping distance. It can be either positive or negative~\cite{boff2002,ge2004}. By defining the MR in $\%$ as $100\times(R(H,T)/R(0,T)-1)$, from equation~\ref{equation_2} one finds:

\begin{equation}
MR[\%]=100~(\frac{exp(AH^2)}{1+<P_iP_j>m^2}-1)
\label{equation_3}\end{equation}

\begin{figure}\vspace{-18cm}\center{\includegraphics[width=22cm]{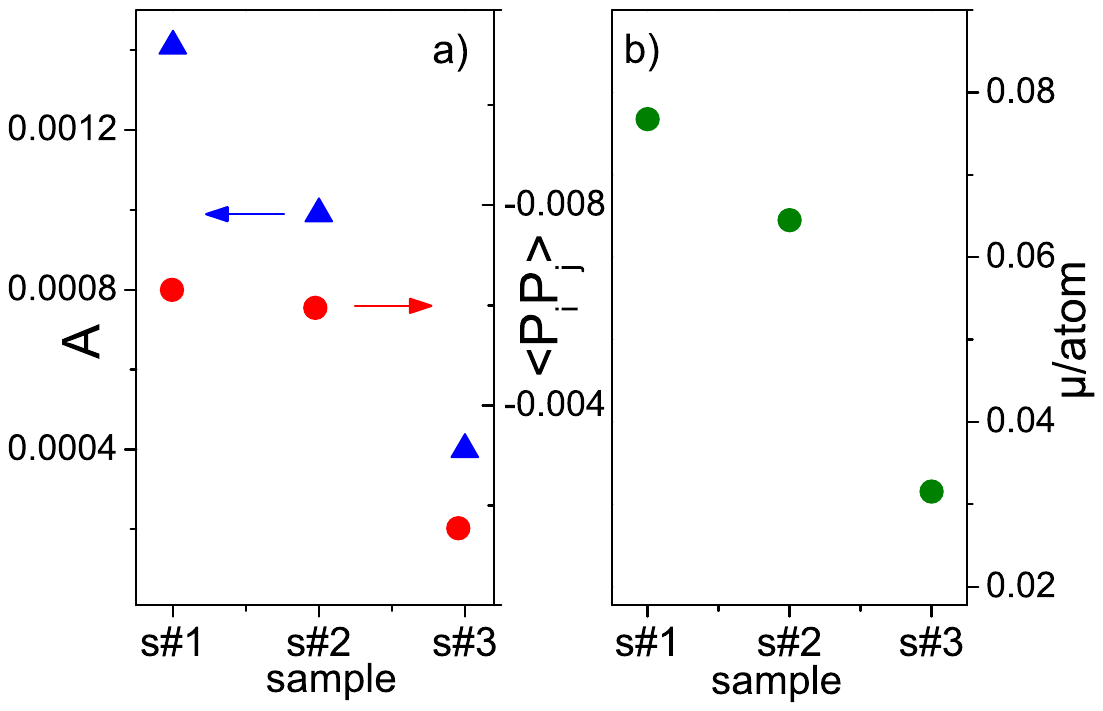}}
\caption{Fit parameters of equation~\ref{equation_3} for MR data taken at 2~K for samples \#1, \#2 and \#3. a) The parameter $A$ and the correlation function $<P_iP_j>$ decrease by increase of the irradiation time, i.e. for more metallic samples, see text for details. b) Magnetic moment $\mu$ for the three samples. The value of $\mu$ is calculated by considering a grain diameter of 3.1 (262 atoms), 3.4 (347) and 4 (564), for samples \#1, \#2 and \#3, respectively in accordance with TEM investigations~\cite{deteresa} and the analysis of the irradiation experiments~\cite{sachser}. }
\label{parameters}\end{figure}

\begin{figure}\vspace{-18cm}\center{\includegraphics[width=22cm]{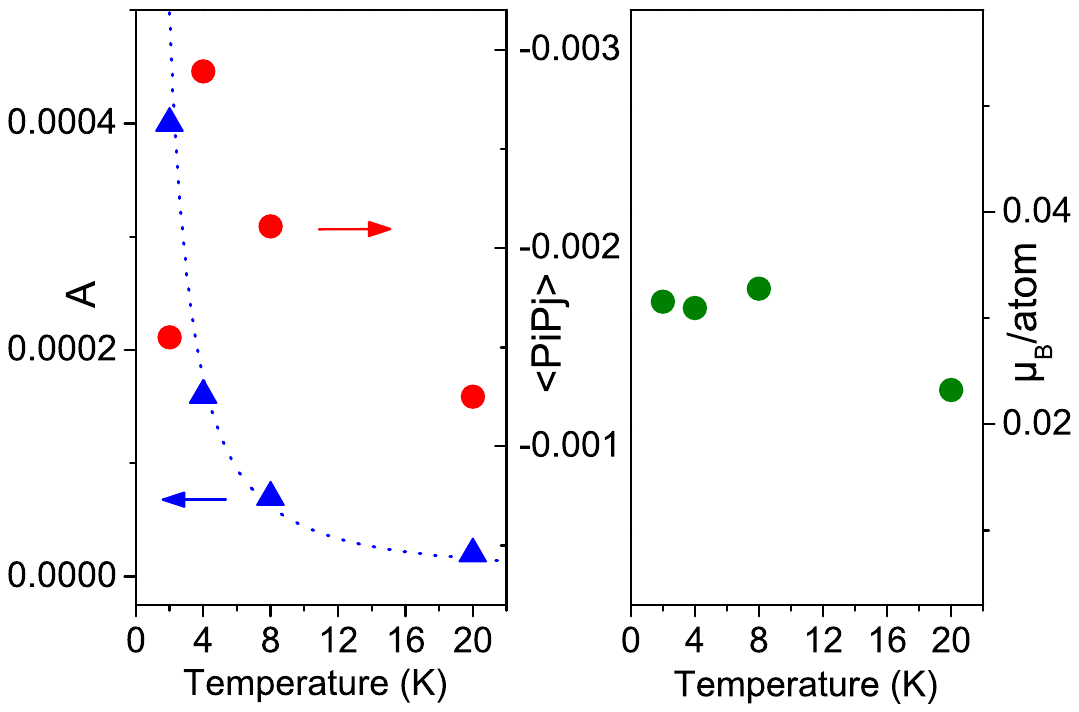}}
\caption{Temperature dependence of the fit parameters for sample~\#3. Left: Parameters $A$ and correlation function $<P_iP_j>$. The dotted-line indicates the $A\sim T^{-3/2}$ expected behavior for ES-VRH. Right: Temperature dependence of the magnetic moment in $\mu_B$/atom.}
\label{parameters_temperature}\end{figure}

In Fig.~\ref{fit_MR} we plot the MR data of samples $\#$1 and $\#$3 taken at 2~K. The data show two inflection points at about 0.3~T and 1.0~T, respectively. A satisfactory fit can be obtained for $P_{i,j}=0$ and $m=0$ for H~$\leq$~0.3~T, see red-dashed line of Fig.~\ref{fit_MR}. However, for larger fields the fit deviates significantly from the experimental data. At large fields in the frame of the ES theory~\cite{shklovskii} one expects a $R(H)\sim exp(AH^{1/3})$ dependence of the resistance~\cite{schoepe,rosenbaum}. However, the inflection point at $H=1.0$~T cannot be explained within the simple ES theory. The best fit to the data is obtained if we assume a non-zero spin polarization, i.e., $P_{i,j}\neq0$ and $m\neq0$, see blue line in Fig.~\ref{fit_MR}. The fits are made with the support of a gaussian weight function in order to reduce the importance of data points taken at high fields which affect the quality of the fit. This procedure is necessary and allowed because the wave function shrinkage model according to equation \ref{equation_3} is not valid at high fields. The results of the fits made at 2~K for the three samples are reported in Fig.~\ref{parameters}. The three parameters $A$, $<P_iP_j>$ and $\mu$ decrease with increasing irradiation dose, i.e., increasing intergrain coupling strength. The parameter $A$ decreases in parallel with the reduction of the MR. The correlation function $<P_iP_j>$ varies according to the modification of the electronic DOS, which depends on the size of the grains~\cite{lammers}. The reduction of the magnetic moment $\mu$ is associated with the increase of the grain size, as found for Fe, Co, Ni clusters~\cite{billas94} and more recently in Pt clusters and nanoparticles~\cite{liu,yamamoto2003}. In Fig.~\ref{parameters_temperature} we report the temperature dependence of the three fit parameters, exemplarily for sample~$\#$3. The parameter $A$ decreases with increasing  temperature, in agreement with the reduction of the MR. The decrease follows a $A\propto T^{-3/2}$ behavior, as expected for ES-VRH.
The temperature dependence of the correlation function $<P_iP_j>$ relates to the changes of the DOS, which was shown to be temperature dependent for Pt metals~\cite{povzner}.
Finally, the magnetic moment $\mu$ is constant till 8~K and shows a decrease at 20~K. Note that the magnetic moments of Fe and Co bulk and clusters were found to be constant till room temperature, while the magnetic moments of Ni clusters show s decrease at lower temperature~\cite{billas94}.

\newpage

\begin{center}
\large\textbf{4. Discussion}\normalsize
\end{center}

The magnetoresistance of magnetic nanostructures prepared by FEBID with precursor gases such as $Fe(CO)_5$ and $Co_2(CO)_8$, has been recently investigated~\cite{fernandez,fab2011Fe}. Owing to the dissociation behavior of these precursors, such nanostructures have a high metal content and their electrical transport behavior is metallic. Differently, deposits obtained from the $(CH_3)_3CH_3C_5H_4Pt$  precursor have a metal content below the percolation threshold. In this case the MR is more difficult to detect because of the low signal at low temperatures. In the present work, our interest lies in studying the MR effect close to the MIT, which we could reach by tuning the electrical conductivity of the Pt-C samples by means of a postgrowth low energy electron irradiation treatment.

Due to the complex interplay of correlation effects, finite size and disorder, the electron transport of granular metals close to the MIT is very hard to treat theoretically~\cite{beloborodov}.
In the weak coupling regime ($g<<1$), sequential tunneling controls the transport at high temperature (thermally activated Arrhenius behavior), while co-tunneling dominates at low temperature (VRH regime). The hopping length $r^*$, which is of the order of the grain size at high temperature, increases by decreasing the temperature, i.e., by moving from sequential tunneling towards inelastic and, then, elastic co-tunneling~\cite{beloborodov}. Moreover, by increasing the intergrain coupling $g$ in the co-tunneling regime towards the MIT, $r^*$ increases to values spanning several grain diameters. In the strong coupling regime ($g>>1$), at high temperature the electron transport, which is coherent only at the scale of the grain size, is dominated by the granular structure. By lowering the temperature a large-scale coherent transport regime develops, the so called granular Fermi liquid~\cite{beloborodov2004}. In this regime the coherence length increases with decreasing the temperature and increasing the intergrain coupling strength. The hopping length, on the insulating side of the MIT, and the coherence length, on the metallic side of the MIT, characterize the length-scale for which the phase of the electron wave function is coherent. As a consequence, the theoretical descriptions available for the weak and strong coupling regimes are expected to merge in a new, not yet available, description valid around the MIT characterized by coherent transport behavior. Interestingly, our experimental results presented in this work indicate that both, properties characteristic of the strong- and of the weak-coupling regimes are present in Pt-C samples close to the MIT. In fact, granular metals in the strong coupling regime ($g\gg1$) show a transition from $\sigma\sim \sqrt{T}$ to $\sigma\sim lnT$ behavior with increasing temperature. The same behavior is found in our deposits, see Figs.~\ref{conductivity} and \ref{conductivity_2}, for $0.25< g<0.5$\cite{sachser}, i.e., close to the MIT. Therefore, for disordered nanogranular Pt-C FEBID deposits the same electrical transport behavior as theoretically predicted for ordered granular lattices in the strong-coupling regime are already found for intermediated coupling strength $0.25< g<0.5$.

In the presence of a magnetic field the nanostructures investigated in this work show a pronounced positive MR. As shown in the previous section, we are able to get satisfactory fits of the data based on the wave-function shrinkage model extended to magnetic nanograins. With an estimated coupling strength of $0.25< g<0.5$ these fits are successful in spite of the fact that we use the theoretical approach outside of its immediate validity range ($g\ll$1).

An important observation is that any fit attempt with non-magnetic nanograins is only satisfactory in a small field range, for $H\leq0.3$~T, but cannot explain the double inflection point present in the measurements. If one considers the Pt nanograins as magnetic, i.e., $P_{i,j}\neq0$ and $m\neq0$, see Fig.~\ref{fit_MR}, the fit is satisfactory up to $H\simeq2.5$~T, which includes both inflection points. Now we turn to discuss the possible reasons for which the Pt nanograins are magnetic. The electronic band structure of Pt bulk is characterized by a DOS at the Fermi level close to satisfy the Stoner criterion for ferromagnetism. Therefore small changes of the band structure can lead to ferromagnetism as found in Au, Pd and Pt nanoparticles~\cite{yamamoto2003,shinohara,yamamoto2004,liu}. An increase of the DOS at the Fermi level due to confinement effects has been theoretically predicted for thin films and clusters~\cite{blügel-watari}. Surface-effects play a role also in nanoparticles since the number of atoms at the surface is comparable with the one in the core. This effect is expected to decrease by increasing the particle size, as it is found in our investigation, see Fig.~\ref{parameters}. A source of local magnetic moments can be induced by the enhancement of the DOS in proximity of twin boundaries, as found in Au and Pt clusters~\cite{sampedro,garcia}. However, for our case we cannot discard this hypothesis since TEM investigations carried out on Pt-based FEBID nanostructures in the as-grown or post-irradiated state~\cite{deteresa,botman2009} are scarce. Nevertheless, they do not show such twinned structures. A further possible source of induced spin polarization are magnetic impurities. EDX measurements exclude the presence of magnetic impurities in the few percent range in our samples. Smaller contaminations in the SEM, not detectable by EDX, cannot be completely ruled out since the FEBID system is also used for preparing Co nanostructures employing the $Co_2(CO)_8$ precursor. However, care is taken to prevent cross contaminations and based on the base pressure of the system we can estimate the maximum impurity level of $Co_2(CO)_8$ in proximity to the point of Pt-C deposition to be $\approx10^{-4}$. A last possible source of permanent magnetic moments is the interaction effect between the nanograins and the surrounding atoms in the matrix. This effect was first considered in the case of
thiol capped Au nanoparticles~\cite{crespo}. In this case the presence of magnetic moments is associated with the spins of localized holes generated through Au-S bonds. Such induced localized magnetic moments were considered as an additional source to the magnetism of Pt-C-capped nanoparticles prepared by laser ablation~\cite{garcia} and they might also contribute to the magnetism of our samples together with the increase of the DOS at the Fermi level due to the finite-size effect. We consider this to be the most likely explanation.

Pt clusters made of 13 atoms in a zeolite have a magnetic moment $\mu=0.45~\mu_B$~\cite{liu}. Pt nanoparticles with diameter ranging from 3.8~nm to 2.3~nm in polymer have a magnetic moment ranging from $\mu_B=0.002$ to $\mu_B=0.012~$, respectively~\cite{yamamoto2003,liu}. In our experiment the Pt nanograins have diameter ranging from 4.0~nm to 3.1~nm and we obtain a magnetic moment from the MR fits ranging from $\mu_B=0.032$ to $\mu_B=0.077~$, respectively, see Fig.~\ref{parameters}b. In both cases the magnetic moment decrease with increasing particle diameter. Interestingly, the fit of our data gives values of $\mu$ of about one order of magnitude larger than those obtained in Ref.~\cite{yamamoto2003,liu}, which we tentatively attribute to the different matrix surrounding the Pt nanoparticles.

The parameter $A$ decreases with temperature in agreement with the ES-VRH modeling, as written above. Furthermore, $A$ decreases with increasing irradiation time, i.e., for stronger coupling. In particular, the increase of the localization length $\xi$ is overcompensated by the decrease of the characteristic temperature $T_0$. This was shown for amorphous carbon films in the ES-VRH regime, where $T_0$ could be directly extracted from the experiment~\cite{wang2013}.

In the following we discuss some energy scale aspects relevant for the interpretation of the experimental results obtained in the present work. First we direct attention to the applicability of the wave function shrinkage model, which was originally developed for Mott-VRH. In the VRH regime at low temperature the hopping of an electron from grain to grain is blocked due to the presence of the Coulomb gap $\Delta$ in the electron excitation spectrum. At very small tunneling conductances, i.e., for Mott-VRH, the Coulomb gap is equal to the charging energy $E_c$. With increasing coupling $\Delta$ is reduced due to virtual electron tunneling to neighboring grains. At even larger couplings the Coulomb gap becomes exponentially suppressed, $\Delta\approx gE_c exp(-\pi zg)$, with $z$ being the coordination number of the grains~\cite{beloborodov2005}. By introducing the values corresponding to our samples in the previous equation, i.e., $g\approx0.2$, $z\simeq12$ for an assumed close packing of the grains and $E_c/k_B\approx430$~K, the value of the Coulomb gap is $\Delta<1$~K. Therefore, in the temperature regime investigated in our work, the correlation effects are strongly reduced so that the wave function shrinkage model, valid for Mott-VRH, may be also applied in the limit of the correlated-VRH regime. Within this frame it has to be noticed that in Ref~\cite{schoepe} numerical calculations were carried out, which extend the wave function shrinkage model to the ES-VRH regime. Another important quantity characterizing the granular system is the mean energy level spacing $\delta$ of the grains, which for our samples is of the order of $\delta=\mathcal{O}({100}~\mu eV$)~\cite{sachser}. The level splitting, together with the Coulomb gap, determine the fine structure of the DOS at the Fermi level. From our experimental data we deduce that some electronic states are spin polarized and contribute to tunneling, since they are placed at the Fermi level. The energy scale of the fine structure can be compared with the relative shift of the DOS of two neighboring grains due to the applied bias voltage used for the electrical transport measurements. By considering grains with perimetrical separation of 4~nm, a voltage drop of 6~mV between the contact electrodes  and a length of 5~$\mu$m for the sample, we obtain a voltage drop between two neighboring Pt grain $\Delta V\approx5~\mu V$. Moreover, since in the co-tunneling regime more grains are involved in the electron transport, the effective voltage drop can increase up to the same order of magnitude of the value of the level splitting. Therefore, the relative shift of the DOS of the grains involved in the transport process is significant and may lead to an inversion of the spin polarization, i.e., $<P_iP_j> <0$. If the inversion takes place or not, depends on the details of the spin-polarized DOS of the surface states, on the electronic structure and on the thickness of the tunneling barrier~\cite{deteresa99}. For the samples investigated in this work the correlation function $<P_iP_j>$ is always negative and it decreases by increasing intergrain coupling strength. Note that a similar trend has been observed in magnetic granular metals with $Co$ or $Fe$ nanograins embedded in a crystalline insulating $Cr_2O_3$ matrix~\cite{ge}. In particular, a change from positive to negative values of $<P_iP_j>$ was measured by increasing the volume fraction of the magnetic grains.

\newpage

\begin{center}
\large\textbf{5. Conclusions}\normalsize
\end{center}

In this work we have investigated the electrical and the magneto-transport properties of Pt-C granular metals fabricated by FEBID. The measurements were carried out for samples carefully tuned to be close to the metal-insulator-transition, for which no theoretical treatment has been suggested so far.

The result is twofold: On the one hand, the conductivity temperature dependence shows a transition from $\sigma\sim \sqrt{T}$ to $\sigma\sim lnT$ with increasing temperature, as expected in the strong-coupling regime~\cite{beloborodov2003,sachser}. On the other hand, the magnetoresistance is described within the wave-function shrinkage model, originally derived for disordered systems in the weak-coupling regime.

We find that the magnetoresistance is positive and its value rises at low temperatures and from the metallic side towards the MIT . In order to fit the experimental data spin-dependent tunneling has to be included in the modeling. As a consequence, we are led to consider the Pt nanograins as magnetic. The values of the MR, of the magnetic moment $\mu$ as well as their temperature and grain size dependence are in agreement with the expected trends known from the literature for other granular metals with magnetic grains. Our findings suggest that the wave-function shrinkage model is able to describe the MR of granular metals close to the MIT.

\begin{center}
\large\textbf{Acknowledgments}\normalsize
\end{center}

The authors acknowledge financial support by the Beilstein-Institut, Frankfurt/Main, Germany, within the research collaboration \emph{NanoBiC}.

\newpage

\end{document}